\documentclass[12pt]{article}

\usepackage{amsmath,amssymb,a4wide,ulem}
\usepackage{amscd}
\usepackage{amsfonts}
\usepackage{mathrsfs}
\usepackage{fancybox} 
\usepackage{enumerate}
\usepackage{accents} 

\usepackage{graphicx}
\usepackage[dvips]{color}

\def\tr{\mathrm{tr}}

\def\det{\mathrm{det}}

\def\Im{\mathrm{Im}}
\def\Re{\mathrm{Re}}

\def\NO#1{\text{:}#1\text{:}}

\newcommand{\wt}{\widetilde}

\def\={\stackrel{\bullet}{=}}

\def\({\left(}
\def\){\right)}
\def\[{\left[}
\def\]{\right]}

\def\cR{{\cal R}}

\def \be {\begin{equation}}
\def \ee {\end{equation}}
\def \beqa {\begin{eqnarray}}
\def \eeqa {\end{eqnarray}}
\def \beal#1 {\begin{align}#1\end{align}}
\def \bes#1 {\begin{equation}\begin{split}#1\end{split}\end{equation}}
\def \nn {\notag\\}

\def\bra#1{\langle #1 |}

\def\ket#1{|#1 \rangle}

\def\aver#1{\left\langle #1 \right\rangle}

\begin{document}


\begin{center}
{\Large\bf Special flow equation and GKP-Witten relation
}
\end{center}
\bigskip
\bigskip
\begin{center}
Sinya Aoki$^1$, Janos Balog$^2$, Tetsuya Onogi$^3$ and Shuichi Yokoyama$^1$
\end{center}
\bigskip
\bigskip

\begin{center}
$^1$ {\it Center for Gravitational Physics and Quantum Information,\\
Yukawa Institute for Theoretical Physics, Kyoto University,\\
Kitashirakawa-Oiwakecho, Sakyo-Ku, Kyoto 606-8502, Japan}

\bigskip
$^2$ {\it Institute for Particle and Nuclear Physics,\\ 
Wigner Research Centre for Physics,\\ 
H-1525 Budapest 114, P.O.B. 49, Hungary} 

\bigskip
$^3$ {\it Department of Physics, Osaka University,\\ 
Toyonaka, Osaka 560-0043, Japan}
\end{center}


\bigskip
\bigskip
\bigskip
\bigskip

\begin{abstract}
{We develop a framework for the reconstruction of the bulk theory dual to conformal field theory (CFT) without any assumption by means of a flow equation. To this end we investigate a minimal extension of the free flow equation and find that at a special parametrization the conformal transformation
for a normalized smeared operator exactly becomes the isometry of anti-de Sitter space (AdS).
By employing this special flow equation to O$(N)$ vector models, we explicitly show that the AdS geometry as well as the scalar field satisfying the GKP-Witten relation concurrently emerge in this framework.}
\end{abstract}

\bigskip
\bigskip

\begin{center}
\end{center}

\newpage

\section{Introduction}

The AdS/CFT duality plays a key role in understanding the holographic aspect of gravitation and may
give a hint for quantum gravity \cite{Maldacena:1997re}. 
Although this duality is supposed to originate in the open/closed string duality and can be tested by explicit calculations \cite{Gubser:1998bc,Witten:1998qj}, the full understanding of the duality including the mechanism of the emergence of the additional direction has not been attained yet. (See \cite{Heemskerk:2010hk} for instance.) 

While several approaches for the bulk reconstruction from boundary CFT have already 
developed considerably (see \cite{Aharony:1999ti,DHoker:2002nbb,Skenderis:2002wp,Benna:2008yg,Harlow:2018fse,Kajuri:2020vxf} for reviews and references therein), 
they assume the existence of an AdS spacetime a priori and only succeeded in reproducing excitations around the AdS background.
The present authors have explored an alternative approach 
employing a course-graining technique called flow equation \cite{Aoki:2015dla,Aoki:2016ohw,Aoki:2016env,Aoki:2017bru,Aoki:2017uce,Aoki:2018dmc,Aoki:2020ztd} 
to realize an emergent bulk AdS geometry itself from CFT. 
A flow equation
generally describes a non-local course-graining of the elementary fields of a given quantum field theory so as to remove contact divergences. This is different from 
 the standard approach, in which the course-graining is applied to
singlet or gauge invariant operators. 
Due to this property the flow equation enables one to construct a class of composite operators which behave as geometric objects in the holographic space after taking the quantum average. 
In particular, taking the vacuum expectation value (VEV) of the metric operator yields a symmetric tensor in a one dimension higher space, which has the interpretation as a certain information metric \cite{Aoki:2017bru}. This is an advantage in this framework, which allows the reconstruction of the bulk theory without assuming the bulk geometry from the beginning. 

The study by means of the flow equation so far has been focused only on the bulk reconstruction for the vacuum state in CFT. 
The next step is to extend this reconstruction to excited states. (See also \cite{Kabat:2011rz,Botta-Cantcheff:2015sav,Kabat:2015swa,Christodoulou:2016nej}.)
The goal of this paper is to develop a framework for it. 
In fact it is possible to continue this program by using the free flow equation as was done previously, but our preliminary studies indicated that our results for excited scalar states using the simplest smearing are difficult to give a geometric interpretation. In particular, 
the  bulk scalar field constructed using the flow equation
does not satisfy the equations of motion on the AdS background. 

In this letter, 
we propose a modified flow equation
and study correlation functions on the emergent bulk spacetime involving excited states by a primary scalar operator. 
We find that some special flow equation maps the conformal symmetry to the bulk AdS isometry at the operator level and allows us to have the expected bulk interpretation for both the background geometry and a scalar field excitation, as we shall see below.

\section{Framework}
\label{sec:Proposal}

\subsection{Interpolating flow}
In this section we develop a framework to construct the bulk counterpart for a primary scalar state in Euclidean CFT by the flow equation approach. 
For a concrete illustration, we consider a $d$-dimensional $O(N)$ vector model with conformal symmetry,
though  we expect that the framework itself and some results can be applied to other CFTs which enjoy the holographic description.
We denote the elementary scalar field in the vector representation of $O(N)$ by $\varphi^a(x)$. The two-point correlation function is 
$\aver{\varphi^a(x_1) \varphi^b(x_2) }= \delta^{ab} {C_0 \over |x_1-x_2|^{2\Delta}}$ where $C_0$ is a constant and $\Delta$ is the conformal dimension. 
In the previous studies of the flow equation approach, we have used the simplest smearing defined by the free flow equation  
\beqa
\partial_\eta \phi^a(x;\eta) =\partial^2 \phi^a(x;\eta), 
\label{eq:freeflow}
\eeqa
with $\phi^a(x;0)=\varphi^a(x)$, where $\eta\geq0$ is the flow parameter, and $\partial^2=\delta^{\mu\nu}\partial_\mu\partial_\nu$. 
However, as mentioned in the introduction, we have encountered a difficulty in the bulk interpretation for an excited state by this simplest smearing. 

Therefore we shift the gear to modify the flow equation suitably for this purpose.  
A minimal extension of the free flow equation \eqref{eq:freeflow} is to  add a second derivative term with respect to $\eta$: 
\be
(-\alpha \eta \partial_\eta^2 +\beta \partial_\eta ) \phi^a(x;\eta) = \partial^2 \phi^a(x;\eta), 
\label{interpolatingFlow}
\ee
where $\alpha$ and $\beta$ are dimensionless real parameters. 
For later convenience, we refer to this flow as the interpolating flow.
We restrict the range of parameters as $\alpha, \beta\geq0$ and it turns out that one can find a desired flow even with this condition.
We solve the equation \eqref{interpolatingFlow} performing the convolution
$
\phi^a(x;\eta) = \int d^dy \rho_\eta(x-y) \varphi^a(y)
\label{convolution}
$, 
with $\rho_0(x)= \delta^d(x)$. 
In the Fourier space the smearing function $\rho_\eta$ satisfies 
\be
(-\alpha \eta \partial_\eta^2 + \beta\partial_\eta ) \rho_\eta(p) = -p^2 \rho_\eta(p), 
\ee
where $ \rho_0(p)=1$. 
This differential equation can be solved by using the modified Bessel function of the second kind as
$
\rho_\eta(p)= {2\over \Gamma(\nu)} \wt p^{\nu} K_{\nu}\left(2\wt p\right)
$, 
where $\wt p =\sqrt{\frac \eta \alpha } p$ with $p=\sqrt{p^2}$ and $\nu = (\alpha+\beta)/\alpha$.
Another independent solution involving $K_{-\nu}\left(2\wt p \right)$ is excluded by the initial condition.
Then the smearing function in the coordinate space is computed as 
\beqa
\rho_\eta(x)
&=& \frac{2}{\Gamma (\nu)  }  {2\pi^{d\over2}\over (x/2)^{\frac{d-2}2}} \int_0^\infty {dpp^{\frac d2}\over (2\pi)^d}  J_{\frac {d-2}2}(px)\wt p^{\nu} K_{\nu}\left(2\wt p\right)\nn 
&=& \frac{\Gamma(\nu+\frac d2)}{\Gamma (\nu)  } \left(\frac \alpha{4\pi\eta}\right)^{\frac{d}2} \left(1+{\alpha x^2 \over 4\eta}\right)^{-(\nu+\frac d2)}, 
\label{rhoeta}
\eeqa
where $J_\mu(x)$ is the Bessel function of the first kind and in the second equation we used a formula found in \cite{Gradshteyn:2014aa}
\beal{
\int_0^\infty \!\!\! dp p^{\bar\mu+\bar\nu+1 } J_{\bar\mu}(ap)   K_{\bar\nu}\left(bp\right)={(2a)^{\bar\mu} (2b)^{\bar\nu} \over (a^2+b^2)^{\bar\mu+\bar\nu+1}}\Gamma(\bar\mu+\bar\nu+1) \notag
}
for $\Re( b )> \vert\Im( a )\vert,  \Re(\bar\mu)>\vert\Re(\bar\nu)\vert -1$. 
This implies $\nu<d/2$.
On the other hand, $\rho_\eta(x) \to \delta^d(x)$ in the limit $\eta\to0$ as long as $\nu>0$. Thus the new parameters are restricted so that 
\be 
0<\nu<d/2.
\label{nurange}
\ee 

One can check that the limit of $\alpha\to0$ reduces to the case of the free flow. 
In the limit, $\nu$ goes to infinity, and with the help of the Stirling
formula we find 
\beal{
\lim_{\alpha\to 0} \rho_\eta(x)|_{\beta=1} 
= \left(\frac 1{4\pi\eta}\right)^{\frac d2} e^{-x^2 \over 4\eta}, 
}
which is indeed the smearing function of the free flow (\ref{eq:freeflow}).

\subsection{Bulk information metric} 
\label{sub_metric}

The flow equation method provides a way to compute a bulk (or $d+1$-dimensional) metric in the expected holographic space without assuming holographic geometry \cite{Aoki:2017bru, Aoki:2015dla}. 
This can be done by constructing an operator in CFT called a metric operator such that 
\beqa
\hat g_{MN}(x;\eta) &:=&\ell^2\sum_{a=1}^N {\partial \sigma^a(x;\eta) \over \partial X^M} {\partial \sigma^a(x;\eta) \over \partial X^N}. 
\label{eq:metric}
\eeqa
Here $X^M=(x^\mu,z)$, where $z = \sqrt{\eta/\gamma}$ with $\gamma$ a constant, $\sigma^a$ is a normalized smeared field defined by $\sigma^a(x;\eta)=\frac1{\sqrt{\aver{\phi(x;\eta)^2 }} } \phi^a(x;\eta)$ with $\phi(x;\eta)^2=\sum_{b=1}^N\phi^b(x;\eta)^2$, and $\ell$ is a length scale chosen by hand. 
Then the VEV of the metric operator yields a metric in the expected holographic space: $g_{MN}(X)=\aver{\hat g_{MN}(x;\eta)}$. 

Comments on this construction are in order. 
Firstly this is a well-defined operator since the smearing by the flow equation removes the contact divergence. 
Indeed the two-point function of the smeared field is evaluated at the coincident point as $\aver{\sum_a\phi^a(x;\eta)^2 }
= N C_1 z^{-2\Delta}$, where
\beqa
C_1 =\gamma^{-\Delta} \int d^d y_1 d^d y_2 \rho_1\left(y_1\right) \rho_1\left(y_2\right) \frac{C_0}{|y_{1}-y_2|^{2\Delta}}.
\eeqa
Secondly due to this construction the classical geometry emerges after taking quantum average as well as the large $N$ limit \cite{Aoki:2015dla}. This can be seen from the behavior of the VEV of the Einstein tensor operator $G_{MN}(\hat g_{PQ})$ \cite{Aoki:2018dmc} 
\beqa
\langle G_{MN} (\hat g_{PQ}) \rangle &=& G_{MN} ( \langle\hat g_{PQ}\rangle ) + O\left({1/ N}\right)
\eeqa
due to the large $N$ factorization. 
The leading term gives the classical AdS value, while the correction term is interpreted as the quantum correction in the bulk.   

Thirdly the induced metric $g_{MN}(X)$ can be interpreted as the Bures information metric for a mixed state defined by $\rho(X) = \sum_{a=1}^N \ket{\sigma^a(x;\eta)}\bra{\sigma^a(x;\eta)}$, where $\ket{\sigma^a(x;\eta)}=\sigma^a(x;\eta)\ket0$ \cite{Aoki:2017bru}. Note that using the normalized field $\sigma^a(x;\eta)$ is important to ensure $\tr\rho=1$.

It was confirmed in the previous study that the induced metric takes the {standard form of} the AdS one in the Poincar\'e patch for the free flow \eqref{eq:freeflow},
{and it becomes the standard one also for the interpolating flow \eqref{interpolatingFlow} by
a proper choice of $\gamma$, which corresponds to
a redefinition of the radial coordinate $z$. Thus,  as shown in \cite{Aoki:2019dim}},
a modification of a flow equation {can be} interpreted as a different choice of {the coordinate system} (or gauge) in the bulk.
{As in the case of the free flow, however,}
a general interpolating flow is not helpful to figure out the bulk interpretation of a CFT excited state.
In the next subsection, {employing a symmetry argument,}
 we search for a special value of $\alpha$ and $\beta$ by which the bulk interpretation becomes transparent.

\subsection{Special flow} 

An infinitesimal conformal transformation for a primary scalar field $\varphi^a(x)$ is
\beal{
\delta^{\rm conf} \varphi^a(x) =& - {\delta x}^\mu\partial_\mu \varphi^a(x)-  {\Delta \over d}  (\partial_\nu \delta x^\nu) \varphi^a(x),
}
where $\delta x^\mu = a^\mu + \omega^\mu{}_\nu x^\nu + \lambda x^\mu + b^\mu x^2 -2 x^\mu (b\cdot x)$.
Then the conformal transformation for the normalized smeared field is given by 
\beal{
\delta^{\rm conf} \sigma^a(x;\eta)
=&  \frac1{\sqrt{\aver{\phi(x;\eta)^2 }} }\int d^d y\rho_\eta(x-y) \delta^{\rm conf} \varphi^a(y) \nn
=& \delta^{\rm diff} \sigma^a(x;\eta) +\delta^{{\rm extra}}\sigma^a(x;\eta), 
}
where we decomposed it into the $d+1$-dimensional AdS isometry $\delta^{\rm diff} \sigma^a$ written as 
\beal{
\delta^{\rm diff} \sigma^a(x;\eta) =& - {\bar\delta x}^\mu\partial_\mu \sigma^a(x;\eta) - \bar\delta z\partial_z \sigma^a(x;\eta) 
\label{diff}
}
with
$\bar\delta x^\mu =\delta x^\mu + \frac {4\gamma}\alpha z^2 b^\mu$, 
$\bar\delta z = (\lambda - 2b\cdot x)z$, 
and the rest 
\beqa
\delta^{{\rm extra}} \sigma^a(x;\eta)
= {2(\nu -\frac d2 + \Delta)\over \sqrt{\aver{\phi(x;\eta)^2 }}} 
\int d^dy\, b\cdot y \rho_\eta(y)  \varphi^a(x-y). \notag
\eeqa
Therefore, $\delta^{{\rm extra}} \sigma^a(x;\eta)$ vanishes for all conformal transformations if and only if $\nu =\frac d2 - \Delta$. In this case the conformal transformation directly converts into the general coordinate transformation on $\sigma^a$ in the bulk with an extra coordinate $z$,
which becomes the AdS isometry for $\gamma = \alpha/4$. 
Thus the conformal invariance guarantees that the bulk geometry is
AdS as $g_{MN}(X)= {R_{\rm AdS}^2 \over z^2}\delta_{MN}$ with $R_{\rm AdS}=\ell \sqrt{\frac{\Delta(d-\Delta)}{d+1}}$.
We refer to this flow with the special parametrization as the special flow. 

For the special flow, the smearing function for $\sigma^a$ is given by $\sigma^a(x;\eta) = \int d^dy\, K(X,y)\varphi^a(y)$, where 
\beqa
K(X,y)=
{\rho_\eta(x-y)  \over \sqrt{\aver{\phi(x;\eta)^2 }}} =C_2 \left({z\over z^2+(x-y)^2}\right)^{d-\Delta}
\label{eq:h_kernel}
\eeqa
with 
$C_2=\frac{\Gamma(d-\Delta)}{ \Gamma(d/2-\Delta)} \sqrt {(\alpha/4)^{\Delta}\over N \pi^d C_1 }$.
This kernel is in fact formally the same as the boundary to bulk propagator \cite{Witten:1998qj}.
Thus this result in some sense reproduces the standard well-known result of the AdS/CFT correspondence only by symmetry argument, though there is a conceptually important difference between the standard bulk reconstruction and the one by means of flow equations. 
That is, in the former approach the smeared objects are singlet or gauge invariant, while in the latter they are elementary fields of the theory. 
Accordingly, the dimension of a smeared object in the latter approach is less than ${d\over 2}$  following from \eqref{nurange}, which is complementary to the condition in the standard bulk reconstruction that the dimension is greater than $d-1$.  (See \cite{Hamilton:2006az} for instance.)

Then a natural question is how singlet operators in CFT are realized in the bulk to satisfy the equation of motion.  
This question is addressed in the next section. 

\section{Bulk reconstruction for excited states} 
\label{sec:geometry}

In the previous section we confirmed not only that the bulk geometry corresponding to a CFT vacuum emerges as AdS but also that 
the conformal symmetry properly acts as the AdS isometry on the flowed scalar field $\sigma^a$ for the special flow.
In this section we extend this bulk reconstruction to an excited state by a singlet primary scalar operator in CFT. 

\subsection{Symmetry constraint}

A key to this extension is that the specially flowed operator $\sigma(x;\eta)$ is constructed so that the conformal transformation acts on it just as the AdS isometry with $z=\sqrt{4\eta/\alpha}$ as the AdS radial coordinate. 
Therefore a correlation function in terms of the specially flowed field and CFT primary operators is constrained severely by conformal symmetry.  
To see this concretely, we consider the following correlation function
\beqa
\left \langle G_{M_1M_2\cdots M_{n}}[\sigma](X) T_{\nu_1\nu_2\cdots \nu_{p}}(y)\right\rangle, 
\eeqa
where $G_{M_1M_2\cdots M_{n}}[\sigma]$ is a CFT operator constructed with the specially flowed operator $\sigma$ while $T_{\nu_1\nu_2\cdots \nu_{p}}$ is a primary tensor operator in CFT. Note that these operators are not restricted to singlet ones.
Then the conformal symmetry constrains this correlation function such that 
\beal{ 
&\left \langle  G_{M_1M_2\cdots M_{n}}[\sigma](\tilde X)  T_{\mu_1\mu_2\cdots \mu_{p}}(\tilde y)\right\rangle  \nn 
=&{\partial X^{N_1}\over \partial \tilde X^{M_1}}\cdots  {\partial X^{N_{n}}\over \partial \tilde X^{M_{n}}}
\times J(y)^{-\Delta_T} {\partial y^{\nu_1}\over \partial \tilde y^{\mu_1}} 
\cdots {\partial y^{\nu_{p}}\over \partial \tilde y^{\mu_{p}}}\nn
&\times \left \langle  G_{N_1N_2\cdots N_{n}}[\sigma](X)  T_{\nu_1\nu_2\cdots \nu_{p}}(y)\right\rangle,
\label{eq:constraint}
}
where $\wt y, \wt X$ are a finite conformal transformation for $y$, an AdS isometric one for $X$, respectively, $J(y)=|\det(\partial_\nu \wt y^\mu)|^{1/d}$, $\Delta_T$ is the conformal dimension of $T$.
A generalization to $k$ bulk operators and $s$ boundary operators is straightforward.
This class of correlation functions is systematically studied in the context of the AdS/CFT correspondence
\cite{Papadimitriou:2004ap,Costa:2011dw,Costa:2011mg,Aharony:2020omh}.  

As a first application, let us compute the correlation function of the smeared operator and the unsmeared one. 
In order to satisfy the constraint \eqref{eq:constraint}, the position dependence is completely fixed as 
\be 
\aver{\sigma^a(x;\eta) \varphi^b(y)} = c \delta^{ab}\left(\frac z{(x-y)^2+z^2}\right)^{\Delta}, 
\label{2ptbulkboundary}
\ee
where the constant $c$ is fixed  by rewriting this in terms of the unnormalized smeared operator as   
\be 
\aver{\phi^a(x;\eta) \varphi^b(y)} = \delta^{ab}\frac {c\sqrt{NC_1}}{((x-y)^2+z^2)^{\Delta}}. 
\ee
Taking the limit $\eta\to0$, the left-hand side reduces to the given two-point correlator of the elementary field if and only if $c= C_0/\sqrt{NC_1}$. 
Thus eqs.\eqref{eq:h_kernel} and \eqref{2ptbulkboundary} give
\beal{
&\aver{\sigma^a(x_1;\eta_1)\sigma^b(x_2;\eta_2)} 
\propto  \nn
&\delta^{ab}\int d^dy  \left[\frac{z_1}{(x_1-y)^2+z_1^2}\right]^{\Delta} \left[\frac{z_2}{(x_2-y)^2+z_2^2}\right]^{d-\Delta}. \notag
} 
Note that this can be written by using a shadow operator of $\varphi^c$ denoted by $(\varphi^c)^\star$, whose conformal dimension is $\Delta^\star=d-\Delta$ as 
\beal{
&\aver{\sigma^a(x_1;\eta_1)\sigma^b(x_2;\eta_2)} 
\propto  \nn
&\sum_{c=1}^N\int d^dy \aver{ \sigma^a(x_1;\eta_1) | \varphi^c(y)}\aver{\varphi^c(y)^\star | \sigma^b(x_2;\eta_2)} 
. \notag 
} 
Taking into account the normalization this can be computed as 
\beqa
&&\aver{\sigma^a(x_1;\eta_1)\sigma^b(x_2;\eta_2)} =\frac{\delta^{ab}}N F_{12}(X_1,X_2), 
\label{hyperres} \\
&&F_{12}(X_1,X_2):={}_2F_1\left(\frac{\Delta}{2},\frac{d-\Delta}{2};  
\frac{d+1}{2};1-\frac{\cR^2}{4}\right), 
\nonumber
\eeqa
where $\cR=\frac{(x_1-x_2)^2+z_1^2+z_2^2}{z_1z_2}$ is the SO($1,d+1$) invariant ratio. 
This explicitly shows the absence of the contact singularity of the smeared operator. 

In the next subsection we shall see that this technique is also useful for the bulk reconstruction for excited states. 

\subsection{GKP-Witten relation}

In the framework using flow equations, excitations in the bulk are reconstructed from the corresponding excited state via pre-geometric operators \cite{Aoki:2018dmc}. 
To see this concretely, let us consider a state excited by a primary scalar singlet operator denoted by $S(y)$ and its dimension by $\Delta_S$. 
(In O$(N)$ vector models, $S(y) \propto \sum_a \NO{\varphi^a\varphi^a}(y)$.)
Corresponding to the boundary excitation, the bulk scalar field is expected to be excited. 
If we denote it by $\chi(X)$, the excitation is described by 
$\chi(X) = 
\sum_a
\bra0 \sigma^a(x;\eta)^2 S(y)\ket0. 
$
Let us prove that the field $\chi(X)$ satisfies the equation of motion on the emergent AdS background. This can be seen by determining the position dependence 
similar to \eqref{2ptbulkboundary} implied by the symmetry constraint \eqref{eq:constraint}: 
\be 
\chi(X)
= C_S \left( {z\over (x-y)^2+z^2}\right)^{\Delta_S},
 \label{eq:bulk2bdr}
\ee
where the overall constant $C_S$ remains undetermined by the symmetry alone. This can be explicitly seen for the case of the vector model by taking the large $N$ limit, which leads to $\Delta_S=2\Delta_\varphi$ due to  the large $N$ factorization.  
From this expression it is easy to see that $\chi(X)$ satisfies the equation of motion with the expected mass dependence
$(\Box_{\rm AdS} -m_S^2)\chi(X) =0$, 
where ${\Box _{\rm AdS}} := (z^2(\partial_z^2+\partial^2) -(d-1) z\partial_z)$ is the d'Alembertian on the AdS space, ${m_S^2} = \Delta_S(\Delta_S - d)$. 

In addition it is not difficult to reproduce the the GKP-Witten relation \cite{Gubser:1998bc,Witten:1998qj} by coupling the field $\chi(X)$ to a sufficiently small source field $J(y)$ and integrating all over the space: 
\be 
\chi_J(X) = \sum_{a=1}^N\bra0 \sigma^a(x;\eta)^2 \int d^dy J(y) S(y)\ket0. 
\label{chiJdef}
\ee
Then by using \eqref{eq:bulk2bdr} the asymptotic behavior of the field $\chi_J(X)$ at the boundary is given by 
$\chi_J(X) \to z^{\Delta_S} \aver{S(x)}_J, $
up to an overall constant, where $\aver{S(x)}_J = \aver{S(x) e^{\int d^dy J(y)S(y)}} \sim \int d^dy J(y)\aver{S(x) S(y)}$.

\section{Discussion}
\label{sec:Discussion}
We emphasize that there are conceptual differences between the flow method and the standard bulk reconstruction. (See \cite{Terashima:2017gmc} for a different point of view on the standard bulk reconstruction.)
Although both methods employ formally the same smearing function to construct the bulk field from the boundary field, they are in fact complementary to each other.

\noindent
(i) The smearing in the flow method is applied to the elementary non-singlet field $\varphi^a$ in the Euclidean path integral, while the bulk reconstruction  gives a relation between boundary and bulk free local singlet quantum operators with Lorentzian signature.

\noindent
(ii) In the bulk reconstruction, the boundary CFT primary field must have a conformal dimension larger than $d-1$. (This constraint can be loosened but then the formula needs to be modified{\cite{Aoki:2021ekk,bhowmick2017bulk}}.) On the other hand, $\varphi^a$ in the flow equation must have $\Delta \le d/2$.
This range of $\Delta$ may be especially important for the application of the method to the study of the AdS dual of the large $N$ critical (or even free)
$O(N)$ model in 3 dimensions \cite{Klebanov:2002ja}. The conformal weight of the basic $O(N)$ field is $\Delta=1/2$ (in the free case).

\noindent
(iii) The VEV of the metric field  realizes an emergent bulk AdS geometry from the CFT in the flow method. On the other hand, the existence of
a background AdS spacetime is assumed from the beginning in the bulk reconstruction.

The smeared and normalized field $\sigma^a$ generates not only the bulk geometry but also other degrees of freedom as composite singlet fields in the bulk, similarly to QCD where all hadrons are made of quarks.
In the bulk construction by the flow method, the constraint \eqref{eq:constraint} imposed by the symmetry is more fundamental than the metric,
which is secondary and is fixed by the constraint. The symmetry and its constraint hold at the fully quantized level and at all orders in the large $N$ expansion.

It would be interesting to ask how scalar excited state contributions affect the AdS geometry described by the VEV of the metric field. 
As before, we consider the VEV of the metric operator in the presence of a small $J(0)$ as
\beqa
\bar g_{MN}^J(X) \simeq \langle 0 \vert\hat g_{MN}(x;\eta) \vert 0\rangle + J(0) \langle 0 \vert\hat g_{MN}(x;\eta) \vert S\rangle.
\eeqa
The constraint \eqref{eq:constraint} leads to
\beqa
\langle 0 \vert\hat g_{MN}(X) \vert S\rangle &=& {T^{\Delta_{{S}}}\over z^2}\left[ a_1  \delta_{MN}
+ a_2  T_M T_N \right],
\eeqa
where $
T:= {z\over x^2+z^2},\
T_z:= {x^2-z^2\over x^2+z^2}, \ T_\mu := -{2x_\mu z\over x^2+z^2} ,$
and the undetermined constants are fixed for the free $O(N)$ vector model as
$a_1=0, \ a_2={\sqrt{{2\over N}} \frac{\Delta^2 \ell^2\Gamma(d)}{\Gamma({d\over 2}+1)\Gamma({d\over 2})}},$\\ $\Delta:={d-2\over 2} .$
Thus the metric $\bar g_{MN}^J(X)$ describes an asymptotically AdS space.

Since the above metric $\bar g_{MN}^J(X)$ deviates from the pure AdS metric, there must exist a matter energy momentum tensor in the bulk,
which can be evaluated through the Einstein equation as
\beqa
2\kappa T_{MN}^{(S)} &=& {J a_2(d-1)\over R_{\rm AdS}^2}{T^{2\Delta}\over z^2} \left[\left(\Delta -{d\over 2}\right) \delta_{MN} -\Delta T_M T_N\right].\notag
\eeqa
We need to find a bulk matter theory which realizes this metric and the corresponding energy momentum tensor,
analogously to what we have found for a non-relativistic CFT\cite{Aoki:2019dim}, but leave this issue for future studies.

The constraint  \eqref{eq:constraint} reproduces  bulk to boundary 2-pt functions in the AdS/CFT correspondence
not only for a scalar field but also for an arbitrary bulk field such as a tensor. 
However, 
there exists an issue on the bulk to bulk  2-pt functions constructed with this method.
For example, let us compute the 2-pt function of the singlet scalar $S(x;z)$:
$\langle S(x_1;z_1) S(x_2;z_2) \rangle \nn
= 1+ {2\over N} F^2_{12}(X_1,X_2) + \langle S(x_1;z_1) S(x_2;z_2)\rangle_c,$
where 
the connected part is evaluated as
\beqa
\langle S(x_1;z_1) S(x_2;z_2)\rangle_c = \left(\prod_{i=1}^4\int d^d y_i\right)  K(X_1,y_1)K(X_1,y_2)\nn
K(X_2,y_3) K(X_2,y_4)
\langle \varphi^a(y_1) \varphi^a(y_2) \varphi^b(y_3) \varphi^b(y_4)\rangle_c .  \nonumber 
\label{eq:connected}
\eeqa
If the boundary theory is free, the bulk to bulk 2-pt function has no singularity at $X_1=X_2$. 
It would be interesting to evaluate {the connected part} explicitly in an interacting theory such as the $O(N)$ vector model.
It remains to be seen that in the interacting case a singularity emerges, as expected in a local bulk theory.

\section*{Acknowledgement}
This work is supported in part by the Grant-in-Aid of the Japanese Ministry of Education, Sciences and Technology, Sports and Culture (MEXT) for Scientific Research (Nos.~JP16H03978,  JP18K03620, JP18H05236, JP19K03847, JP22H00129), and by the NKFIH grant K134946. 
J.B. acknowledges support from the International Research Unit of Quantum Information (QIU) of Kyoto University Research Coordination Alliance, and also would like to thank the Yukawa Institute for Theoretical Physics at Kyoto University, where most of this work has been carried out, for support and hospitality by the long term visitor program.
T. O. would like to thank YITP for their kind hospitality during his stay for the sabbatical leave from his home institute.

\bibliographystyle{unsrt2}
\bibliography{CFT2AdS}

\end{document}